\documentstyle[prd,preprint,aps]{revtex}
\begin{document}
%\draft command makes pacs numbers print
\draft
\title{Production of the MSSM Higgs Bosons at\\
       Next Generation Linear $e^+e^-$ Colliders}

\author{A. Guti\'errez-Rodr\'{\i}guez $^{1}$, M. A. Hern\'andez-Ru\'{\i}z $^{1}$ and O. A. Sampayo $^{2}$}

\address{(1) Facultad de F\'{\i}sica, Universidad Aut\'onoma de Zacatecas\\
          Apartado Postal C-580, 98060 Zacatecas, Zacatecas M\'exico.}

\address{(2) Departamento de F\'{\i}sica, Universidad Nacional del Mar del Plata\\
             Funes 3350, (7600) Mar del Plata,  Argentina.}

\date{\today}
\maketitle
\begin{abstract}
% insert abstract here
We study the production of the Higgs bosons predicted in the Minimal Supersymmetric
extension of the Standard Model $(h^0, H^0, A^0, H^\pm)$, with the reactions
$e^{+}e^{-}\rightarrow b\bar b h^0 (H^0, A^0)$, and
$e^+e^-\to \tau^-\bar \nu_\tau H^+, \tau^+\nu_\tau H^-$, using the helicity
formalism. We evaluate cross section of $h^0, H^0, A^0$ and $H^\pm$ in the
limit when $\tan\beta$ is large. The numerical computation is done considering
two stages of a possible Next Linear $e^{+}e^{-}$ Collider: the first with
$\sqrt{s}=500$ $GeV$ and design luminosity 50 $fb^{-1}$, and the second with
$\sqrt{s}=1$ $TeV$ and luminosity 100-200 $fb^{-1}$.
\end{abstract}

\narrowtext
\section{Introduction}
Higgs bosons \cite{Higgs} play an important role in the Standard Model (SM) \cite{Weinberg};
they are responsible for generating the masses of all the elementary particles
(leptons, quarks, and gauge bosons). However, the Higgs-boson sector is the
least tested one in the SM. If Higgs bosons are responsible for breaking the
symmetry from $SU(2)_L\times U(1)_Y$ to $U(1)_{EM}$, it is natural to expect
that other Higgs bosons are also involved in breaking other symmetries.
One of the more attractive extensions of the SM is Supersymmetry (SUSY) \cite{Nilles},
mainly because of its capacity to solve the naturalness and hierarchy problems
while maintaining the Higgs bosons elementary.

The theoretical frame work of this paper is the Minimal Supersymmetric extension
of the Standard Model (MSSM), which doubles the spectrum of particles of the SM
and the new free parameters obey simple relations. The scalar sector of the MSSM
\cite{Gunion} requires two Higgs doublets, thus the remaining scalar spectrum
contains the  following physical states: two CP-even Higgs scalar ($h^0$ and
$H^0$) with $m_{h^0}\leq m_{H^0}$, one CP-odd Higgs scalar ($A^0$) and a charged
Higgs pair ($H^{\pm}$), whose detection would be a clear signal of new physics.
The Higgs sector is specified at tree-level by fixing two parameters, which can
be chosen as the mass of the pseudoscalar $m_{A^0}$ and the ratio of vacuum
expectation values of the two doublets $\tan \beta = v_{2}/v_{1}$, then the mass
$m_{h^0}$, $m_{H^0}$ and $m_{H^{\pm}}$ and the mixing angle of the neutral
Higgs sector $\alpha$ can be fixed. However, since radiative corrections
produce substantial effects on the predictions of the model \cite{Li}, it is
necessary to specify also the squark masses, which are assumed to be
degenerated. In this paper, we focus on the phenomenology of the neutral
CP-even and CP-odd scalar ($h^0, H^0, A^0$) and charged $(H^\pm)$.  

The search for these scalars has begun at LEP, and current low energy bound
on their masses gives $m_{h^0}$, $m_{A^0}$ $>$ 90 $GeV$ and $m_{H^\pm}$
$>$ 120 $GeV$ for $\tan\beta$ $>$ 1 \cite{Review}.
At $e^{+}e^{-}$ colliders the signals for Higgs bosons are relatively clean
and the opportunities for discovery and detailed study will be excellent. The
most important processes for the production and detection of the neutral and
charged Higgs bosons $h^0$, $H^0$, $A^0$ and $H^\pm$, are: $e^{+}e^{-}\rightarrow
Z^{*}\rightarrow h^0, H^0+Z^0$, $e^{+}e^{-}\rightarrow Z^{*} \rightarrow
h^0, H^0+A^0$, $e^{+}e^{-}\rightarrow \nu \bar \nu + W^{+*}W^{-*}
\rightarrow \nu \bar \nu+h^0, H^0$ (the later is conventionally referred
to as $WW$ fusions), and $e^+e^-\to H^+H^-$ \cite{Komamiya}; precise cross-section
formulas appear in Ref. \cite{Gunion1}. The main decay modes of the neutral Higgs
particles are in general $b\bar b$ decays $(\sim 90 \%)$ and $\tau^+\tau^-$ decays
$(\sim 10 \%)$ which are easy to detect experimentally at $e^+e^-$
colliders \cite{Janot,Carena,Sopczak}. Charged Higgs particles decay predominantly
into $\tau \nu_\tau$ and $t\bar b$ pairs.

In previous studies, the two-body processes of the neutral and charged Higgs
bosons $e^{+}e^{-}\rightarrow h^0(H^0)+Z^0$, $e^{+}e^{-}\rightarrow h^0(H^0)+A^0$
and $e^+e^- \rightarrow H^+H^-$ have been evaluated \cite{Gunion1} extensively.
However, the inclusion of three-body process with heavy fermions $f$, 
$e^{+}e^{-}\rightarrow (f\bar f h^0, f\bar f H^0, f\bar f A^0)$,
\cite{Djouadi1,Cotti} and $e^+e^-\to \tau^-\bar \nu_\tau H^+,
\tau^+ \nu_\tau H^-$ \cite{Gutierrez,Kanemura} at future $e^+e^-$ colliders energies
\cite{NLC,NLC1,JLC} is necessary in order to know its impact on the two-body
mode processes and also to search for new relations that could have a cleaner
signature of the Higgs bosons production.

In the present paper we study the production of SUSY Higgs bosons at $e^{+}e^{-}$ colliders.
We are interested in finding regions that could allow the detection of the SUSY
Higgs bosons for the set parameter space $(m_{A^0}-\tan\beta)$. We shall discuss
the neutral and charged Higgs bosons production $b\bar b h^0 (H^0, A^0)$, and
$\tau^-\bar \nu_\tau H^+$, $\tau^+\nu_\tau H^-$ in the energy range of a future
$e^+e^-$ collider \cite{NLC,NLC1,JLC} for large values of the parameter
$\tan\beta$, where one expects to have a high production. Since the coupling
$h^0b\bar b$ is proportional to $\sin\alpha/\cos\beta$, the cross-section will
receive a large enhancement factor when $\tan\beta$ is large. Similar situation
occurs for $H^0$, whose coupling with $b\bar b$ is proportional to $\cos\alpha/\cos\beta$.
The couplings of $A^0$ with $b\bar b$ and of $H^\pm$ with $\tau^-\bar \nu_\tau, \tau^+\nu_\tau$
are directly proportional to $\tan\beta$, thus the amplitudes will always grow with
$\tan\beta$. We consider the complete set of Feynman diagrams at tree-level
and use the helicity formalism \cite{Howard,Zhan,Werle,Pilkum,Peter,Mangano,Berends}
for the evaluation of the amplitudes. Succinctly, our aim in this work is to
analyze how much the results of the Bjorken Mechanism [Fig. 1, (1.4)] would be
enhanced by the contribution from the diagrams depicted in Figs. 1.1-1.3, 1.5
and 1.6 in which the SUSY Higgs bosons are radiated by a $b (\bar b)$ quark.
For the case of the charged Higgs bosons the two-body mode [Figs. 3.1 and 3.4]
would be enhanced by the contribution from the diagrams depicted in Figs. 3.2,
3.3, and 3.5, in which the charged Higgs boson is radiated by a $\tau^- \bar \nu_\tau$
$(\tau^+ \nu_\tau)$ lepton.

Recently, it has been shown that for large values of $\tan\beta$ the detection
of SUSY Higgs bosons is possible at FNAL and LHC \cite{Lorenzo}. In the papers
cited in Ref. \cite{Lorenzo} the authors calculated the corresponding three-body diagrams
for hadron collisions. They pointed out the importance of a large bottom
Yukawa coupling at hadron colliders and showed that the Tevatron collider
may be a good place for detecting SUSY Higgs bosons. In the case of the
hadron colliders the three-body diagrams come from gluon fusion and this
fact makes the contribution from these diagrams more important, due to the
gluon abundance inside the hadrons. The advantage
for the case of $e^{+} e^{-}$ colliders is that the signals of the processes
are cleaner.

This paper is organized as follows. We present in Sec. II the relevant details
of the calculations. In Sec. III we evaluate cross section for the processes
$e^{+}e^{-}\rightarrow b\bar bh^0 (H^0,A^0)$ and $e^+e^-\to \tau^-\bar \nu_\tau H^+,
\tau^+ \nu_\tau H^-$ at future $e^+e^-$ colliders. Finally, Sec. IV contains
our conclusions.

\section{Helicity Amplitude for Higgs Bosons Production}

When the number of Feynman diagrams is increased, the calculation of the amplitude
is a rather unpleasant task. Some algebraic forms \cite{Hearn} can be used in it to
avoid manual calculation, but sometimes the lengthy printed output from the computer
is overwhelming, and one can hardly find the required results from it. The CALKUL 
collaboration \cite{Causmaecker} suggested the Helicity Amplitude Method (HAM) which can
simplify the calculation remarkably and hence make the manual calculation realistic.

In this section we describe in brief the evaluation of the amplitudes at tree-level, for
$e^{+}e^{-}\rightarrow b\bar b h^0 (H^0, A^0)$ and $e^+e^-\to \tau^-\bar \nu_\tau H^+,
\tau^+ \nu_\tau H^-$   using the HAM \cite{Howard,Zhan,Werle,Pilkum,Peter,Mangano,Berends}.
This method is a powerful technique for computing helicity amplitudes for multiparticle
processes involving massless spin-1/2 and spin-1 particles. Generalization of this method 
that incorporates massive spin-1/2 and spin-1 particles, is given in Ref. \cite{Berends}.
This algebra is easy to program and more efficient than computing the Dirac algebra.

A Higgs boson $h^0, H^0$, $A^0$, and $H^\pm$ can be produced in scattering $e^{+}e^{-}$ via
the following processes:

\begin{eqnarray}
e^{+}e^{-} &\rightarrow& b\bar b h^0,\\
e^{+}e^{-} &\rightarrow& b\bar b H^0,\\
e^{+}e^{-} &\rightarrow& b\bar b A^0,\\         
e^+e^-&\rightarrow& \tau^-\bar \nu_\tau H^+, \tau^+ \nu_\tau H^-.
\end{eqnarray}

The diagrams of Feynman, which contribute at tree-level to the different reaction
mechanisms, are depicted in Figs. 1-3.

However those diagrams with exchange of Higgs bosons instead of gauge bosons
(photon or $Z^0$) have been neglected because of the smallness of the Higgs fermion
coupling. Using the Feynman rules given by the
Minimal Supersymmetric Standard Model (MSSM), as summarized in Ref. \cite{Gunion1},
we can write the amplitudes for these reactions. For the evaluation
of the amplitudes we have used the spinor-helicity technique of Xu, Zhang and
Chang \cite{Zhan} (denoted henceforth by XZC), which is a modification of the
technique developed by the CALKUL collaboration \cite{Causmaecker}. Following
XZC, we introduce a very useful notation for the calculation of the processes
(1)-(4). The complete formulas of the processes (1)-(3) are given in Ref.
\cite {Cotti}, while for the case of process (4), are given in Ref. \cite{Gutierrez}.
We are going to use the same notation and the formulas given in
\cite{Cotti,Gutierrez} and do not reproduce them here.

After writing down the Feynman diagrams corresponding to a given amplitude ${\cal M}$
an usually proceeds to derive an analytic expression for the cross section
$\sum|{\cal M}|^2$, with an appropriate spin and/or color sum or average. The
result, which is usually a function of Minkowsky products of the particle
four-momenta, is then evaluated numerically at phase-space points in the region
of interest.

It is clear that any method which evaluates ${\cal M}$ instead of $\sum|{\cal M}|^2$
will eventually become superior to the standard approach. Indeed, several authors
have stressed this point and proposed alternatives.

We want to argue that it is useful to employ in the evaluation of the amplitude
not vector products like $p_1\cdot p_2$ but rather spinor products like $\bar u(p_1)u(p_2)$.

After the evaluation of the amplitudes of the corresponding diagrams, we obtain
the cross-sections of the analyzed processes for each point of the phase space
using Eqs. (12)-(17), (23)-(28) and (11)-(15) \cite{Cotti,Gutierrez} by a computer
program, which makes use of the subroutine RAMBO (Random Momenta Beautifully Organized)
\cite{Kleiss}. The advantages of this procedure in comparison to the traditional
``trace technique" are discussed in Refs. \cite{Howard,Zhan,Werle,Pilkum,Peter,Mangano,Berends}.

We use the Breit-Wigner propagators for the $Z^{0}$, $h^0$, $H^0$, $A^0$ and
$H^{\pm}$ bosons. For the SM parameters we adopted the following:
$m_b=4.25$ $GeV$, $m_t=175$ $GeV$, $m_\tau =1.78$ $GeV$, $m_\nu =0$, $m_{Z^0}=91.2$ $GeV$,
$\Gamma_{Z^0}=2.4974$ $GeV$, $\sin^2\theta_W=0.232$, which are taken as inputs.
The widths of $h^0$, $H^0$, $A^0$ and $H^{\pm}$ are calculated from the formulas
given in Ref. \cite{Gunion1}. In the next sections we present the numerical
computation of the processes $e^+e^-\to b\bar bh$, $h=h^0, H^0, A^0$ and
$e^+e^-\to \tau^-\bar \nu_{\tau} H^+, \tau^+\nu_\tau H^-$.

\section{Production of the MSSM Higgs Bosons at Next Generation Positron-Electron Colliders}

In this paper, we evaluate total cross section of neutral and charged MSSM Higgs
bosons at next generation $e^{+}e^{-}$ colliders, including three-body mode
diagrams [Figs. 1.1-1.3, 1.5, and 1.6; Figs. 2.1-2.3, 2.5 and 2.6; Figs. 3.2, 3.3,
and 3.5] besides the dominant mode diagram [Fig. 1.4; Fig. 2.4; Figs. 3.1, and 3.4]
consider two stages of a possible Next Linear $e^{+}e^{-}$ Collider: the first
with $\sqrt{s}=500$ $GeV$ and design luminosity 50 $fb^{-1}$, and the second with
$\sqrt{s}=1$ $TeV$ and luminosity 100-200 $fb^{-1}$. We consider the complete set
of Feynman diagrams (Figs. 1-3) at tree-level and utilize the helicity formalism
for the evaluation of their amplitudes. In the next subsections, we present our
results for the case of the different Higgs bosons.

\subsection{Cross section of $h^0$}

In order to illustrate our results on the production of the $h^0$ Higgs boson, 
we present graphs of the cross section as functions of $A^0$ Higgs boson mass
$m_{A^0}$, assuming $m_{t}= 175$ $GeV$, $M_{\stackrel{\sim}t}= 500$ $GeV$ and $\tan\beta > 1$ for
NLC. Our results are displayed in Fig. 4, for the process at three-body
$e^{+}e^{-}\rightarrow b\bar b h^0$ and for the $e^{+}e^{-}\rightarrow (A^0, Z^0)+ h^0$
dominant mode.

The total cross section at colliders energies of 500 $GeV$ and 4 different
values of the fundamental supersymmetry parameter $\tan\beta$, 6, 10, 30, 50
is of the order of 0.1 $pb$. We note from this figure, that the effect of
the reaction $b\bar b h^0$ is light more important that $(A^0, Z^0) + h^0$,
for the interval 80 $GeV$$\leq m_{A^0}\leq 110$ $GeV$ and $\tan\beta =30,  50$.
Nevertheless, there are substantial portions in which the discovery of the
$h^0$ is not possible using either $b\bar b h^0$ or $(A^0, Z^0) + h^0$.

For the case of $\sqrt{s}=1$ $TeV$, the results of the total cross section
of the $h^0$ are shown in Fig. 8. It is clear from this figure that the
contribution of the process $e^{+}e^{-}\rightarrow b\bar b h^0$ becomes
dominant in the interval 80 $GeV$$ \leq m_{A^0} \leq 120$ $GeV$. However,
they could provide important information on the Higgs bosons detection. For
instance, we present in the next section tables that illustrate the events
number for the process $e^+e^-\rightarrow b\bar b h^0$, with $m_{A^0}=100$ $GeV$,
$\sqrt{s}= (500, 1000)$ $GeV$ and $\tan\beta=10,30$.

\subsection{Cross section of $H^0$}

To illustrate our results regarding the detection of the heavy Higgs bosons $H^0$,
we give the total cross section for both processes $e^{+}e^{-}\rightarrow b\bar b H^0$
and $e^{+}e^{-}\rightarrow (A^0, Z^0) + H^0$ in Fig. 5 for $\sqrt{s}= 500$ $GeV$.

The total cross section for this case is of the order of 0.01 $pb$. In this figure,
we observed that the effect of incorporate $b\bar b H^0$ in the detection of the
Higgs boson $H^0$ is more important than the case of two-body mode $(A^0, Z^0)+H^0$,
because $b\bar b H^0$ is dominant in all the interval of 100 $GeV$$\leq m_{A^0}\leq 400$
$GeV$.

For the case of $\sqrt{s}=1$ $TeV$, the results are show in Fig. 9. In this
case the three-body mode $b\bar bH^0$ is light more important that $(A^0, Z^0)+H^0$
for 100 $GeV$$\leq m_{A^0}\leq 450$ $GeV$. In the next section are present 
tables with the events number for $e^+e^- \rightarrow b\bar b H^0$.

\subsection{Cross section of $A^0$}

For the pseudoscalar $A^0$, it is interesting to consider the production mode
in $b\bar b A^0$, since it can have large a cross-section due to the fact that
the coupling of $A^0$ with $b\bar b$ is directly proportional to $\tan\beta$,
thus will always grow with it. In Fig. 6, we present the total cross section
for the process of our interest $b\bar b A^0$, at $\sqrt{s}=500$ $GeV$ and
$\tan\beta =6, 10, 30, 50$.

We note that for $\tan\beta = 6, 10, 30, 50$, both the $b\bar b A^0$ mode and the $(h^0, H^0)+A^0$
mode have cross sections of the order of 0.1 $pb$, and for $m_{A^0}> 175$ $GeV$ the
process $b\bar b A^0$ cover a major region in the parameters space $(m_{A^0}, \tan\beta)$.

On the other hand, if we focus the detection of the $A^0$ at $\sqrt{s}=1$ $TeV$,
the panorama for its detection is more extensive. The Fig. 10 shows the contours
lines in the plane $(\sigma_T-m_{A^0})$, to the production cross section
of $b\bar bA^0$. The contours for this cross section correspond to
$\tan\beta = 6, 10, 30, 50$. The total cross section is of the order of 0.01 $pb$
and the three-body mode $e^+e^-\rightarrow b\bar b A^0$ is more important that
two-body mode $e^+e^-\rightarrow (h^0,H^0)+A^0$ in all the parameters space.

\subsection{Cross section of $H^\pm$}

Our results for the $H^+H^-$ scalars are displayed in Fig. 7, 11 for
the processes at three-body $e^+e^-\to \tau^-\bar \nu_\tau H^+,
\tau^+\nu_\tau H^-$ and for the $e^+e^-\to H^+H^-$ dominant mode.

The total cross section for this reaction with $\sqrt{s}=500$ $GeV$ and
4 different values of $\tan\beta$, 6, 10, 30, 50 are show in Fig. 7. The
cross section is of the order of $\sigma_T \approx 0.1$ $pb$ for $m_{H^\pm}\approx 100$ $GeV$
and $\tan\beta$ large, while for $m_{H^\pm}\approx 200$ $GeV$ is of $\sigma_T \approx 0.01$ $pb$.
We can see from this figure that the effect of the reactions $\tau^-\bar \nu_\tau H^+$
and $\tau^+\nu_\tau H^-$ is slightly more important than $H^+H^-$ for most of
the $(m_{A^0}-\tan\beta)$ parameters space regions. It is precisely in this curve
where the contribution of the processes at three-body is notable. Nevertheless,
there are substantial portions of parameters space in which the discovery of the
$H^\pm$ is not possible using either $H^+H^-$ or $\tau^-\bar\nu_\tau H^+$ and
$\tau^+\nu_\tau H^-$.

For the case of $\sqrt{s}=1$ $TeV$, the results are show in Fig. 11. In both
cases ($\tau^-\bar\nu_\tau H^+$, $\tau^+\nu_\tau H^-$ and $H^+H^-$) the curves
with $\tan\beta$, 6, 10, 30, 50 given $\sigma_T\approx 0.1$ $pb$ for $m_{H^\pm}\approx 175$
$GeV$. While for $m_{H^\pm}\approx 350$ $GeV$ the cross section drop until 0.001 $pb^{-1}$
for $\tan\beta =6$. These cross sections are small, however, the contribution
of the processes at three-body is slightly more important than the process at
two-body. The most important conclusion from this figure is that detection of
the charged Higgs bosons will be possible at future $e^+e^-$ colliders.

\subsection{Total Production of the Higgs Bosons}

We present in Tables I, II our results for the total production of $h^0$, $H^0$,
$A^0$, $H^\pm$ Higgs bosons, taking different values of the center-of-mass energy
$\sqrt{s}$, fundamental supersymmetry parameter $\tan\beta$, luminosity ${\cal L}$,
and the mass of the pseudoscalar $m_{A^0}$. We take the following values representative
of the supersymmetric parameters $m_{A^0}=100$ $GeV$ and $\tan\beta= 10, 30$.
From these results we observed a strong dependence of the supersymmetric parameter
$\tan\beta$, in particularly for $\tan\beta$ large, as well as, of the center-of-mass energy
$\sqrt{s}$, in the production of the different Higgs bosons $(h^0, H^0, A^0, H^\pm)$.
These results make apparent the big importance of investigate the possibility
of detecting the Higgs bosons with the reactions at three-body
$e^{+}e^{-}\rightarrow b\bar b h^0(H^0, A^0)$ and
$e^{+}e^{-}\rightarrow \tau^-\bar \nu_\tau H^+, \tau^+\nu_\tau H^-$, at next
generation linear $e^+e^-$ colliders.

%\vspace*{8mm}
\newpage

\begin{center}
\begin{tabular}{|c|c|c|}
\hline
Total Production of Higgs Bosons&\multicolumn{2}{c|}{${\cal L}$=50 $fb^{-1}$}\\
\cline{2-3}
&$\tan\beta=10$ & $\tan\beta=30$ \\
\hline
\hline
$h^0$           & 1600 & 1800\\
$H^0$           & 700  & 470\\
$A^0$           & 900  & 935\\
$H^+ H^-$       & 7000 & 6500\\
\hline
\end{tabular}
\end{center}

\begin{center}
Table I. Total production of Higgs bosons $h^0, H^0, A^0, H^\pm$ of the MSSM
for $\sqrt{s}=500$ $GeV$ and $m_{A^0}=100$ $GeV$.
\end{center}

\vspace*{5mm}

\begin{center}
\begin{tabular}{|c|c|c|c|c|}
\hline
Total Production of Higgs Bosons&\multicolumn{2}{c|}{${\cal L}$=100 $fb^{-1}$}
         &\multicolumn{2}{c|}{${\cal L}$=200 $fb^{-1}$}\\
\cline{2-5}
&$\tan\beta=10$ & $\tan\beta=30$ & $\tan\beta=10$ & $\tan\beta=30$ \\
\hline
\hline
$h^0$           & 960  & 1150  & 1920  & 2300 \\
$H^0$           & 370  & 220   & 740   & 440 \\
$A^0$           & 560  & 615   & 1120  & 1230 \\
$H^+H^-$        & 4700 & 4900  & 9400  & 9800 \\
\hline
\end{tabular}
\end{center}

\begin{center}
Table II. Total production of Higgs bosons $h^0, H^0, A^0, H^\pm$ of the MSSM
for $\sqrt{s}=1$ $TeV$ and $m_{A^0}=100$ $GeV$.
\end{center}

\section{Conclusions}

In this paper, we have calculated the production of the neutral and charged Higgs
bosons in association with $b$-quarks and with $\tau \nu_\tau$ leptons via the
processes $e^{+}e^{-}\rightarrow b \bar b h$, $h = h^0, H^0, A^0$ and
$e^+e^-\to \tau^-\bar \nu_\tau H^+, \tau^+\nu_\tau H^-$ using the helicity
formalism. We find that these processes could help to detect the possible neutral
and charged Higgs bosons at energies of a possible Next Linear $e^+e^-$ Collider
when $\tan \beta$ is large.

In summary, we conclude that the possibilities of detecting or excluding the 
neutral and charged Higgs bosons $(h^0, H^0, A^0, H^\pm)$ of the Minimal
Supersymmetric Standard Model in the processes $e^{+}e^{-}\rightarrow b\bar b h$,
$h = h^0, H^0, A^0$ and $e^+e^-\to \tau^-\bar \nu_\tau H^+, \tau^+ \bar \nu_\tau H^-$
are important and in some cases are compared favorably with the dominant mode
$e^{+}e^{-}\rightarrow (A^0, Z^0) + h$, $h = h^0, H^0, A^0$ and $e^+e^-\to H^+H^-$
with $\tan \beta$ large. The detection of the Higgs boson will require the use
of a future high energy machine like the Next Linear $e^{+}e^{-}$ Collider.

\hspace{2cm}

\begin{center}
{\bf Acknowledgments}
\end{center}

This work was supported in part by {\it Consejo Nacional de Ciencia y Tecnolog\'{\i}a}
(CONACyT), {\it Sistema Nacional de Investigadores} (SNI) (M\'exico) and
Programa de Mejoramiento al Profesorado (PROMEP). A.G.R. would like to thank
the organizers of the Summer School in Particle Physics and Sixth School on
non-Accelerator Astroparticle Physics 2001, Trieste Italy for their hospitality.
O. A. S. would like to thank CONICET (Argentina).

\newpage

\begin{center}
{\bf FIGURE CAPTIONS}
\end{center}

\vspace{5mm}

\bigskip

\noindent {\bf Fig. 1} Feynman Diagrams at tree-level for $e^{+}e^{-}
\rightarrow b\bar b h^0$. For $e^{+}e^{-}\rightarrow b\bar b H^0$
one has to make only the change $\sin\alpha / \cos\beta \rightarrow
\cos\alpha / \cos\beta$.

\bigskip

\noindent {\bf Fig. 2} Feynman Diagrams at tree-level for $e^{+}e^{-}
\rightarrow b\bar b A^0$.

\bigskip

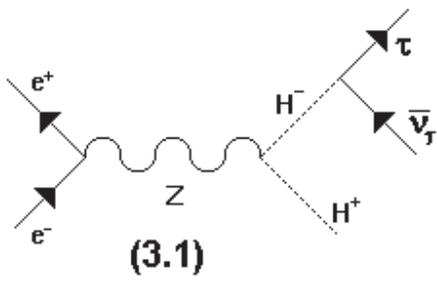
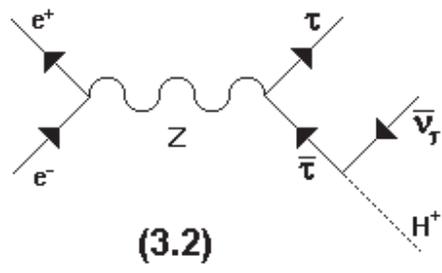
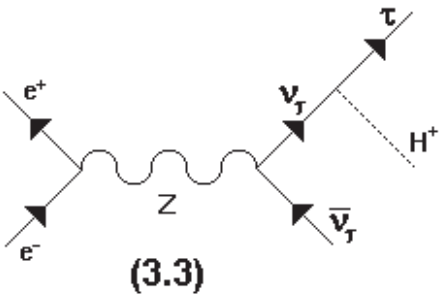
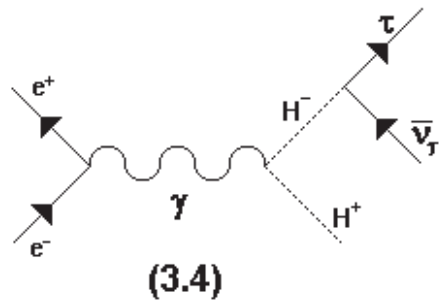
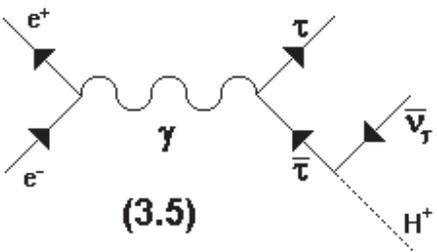
\noindent {\bf Fig. 3} Feynman Diagrams at tree-level for $e^{+}e^{-}
\rightarrow \tau^-\bar \nu_\tau H^+, \tau^+\nu_\tau H^-$.

\bigskip

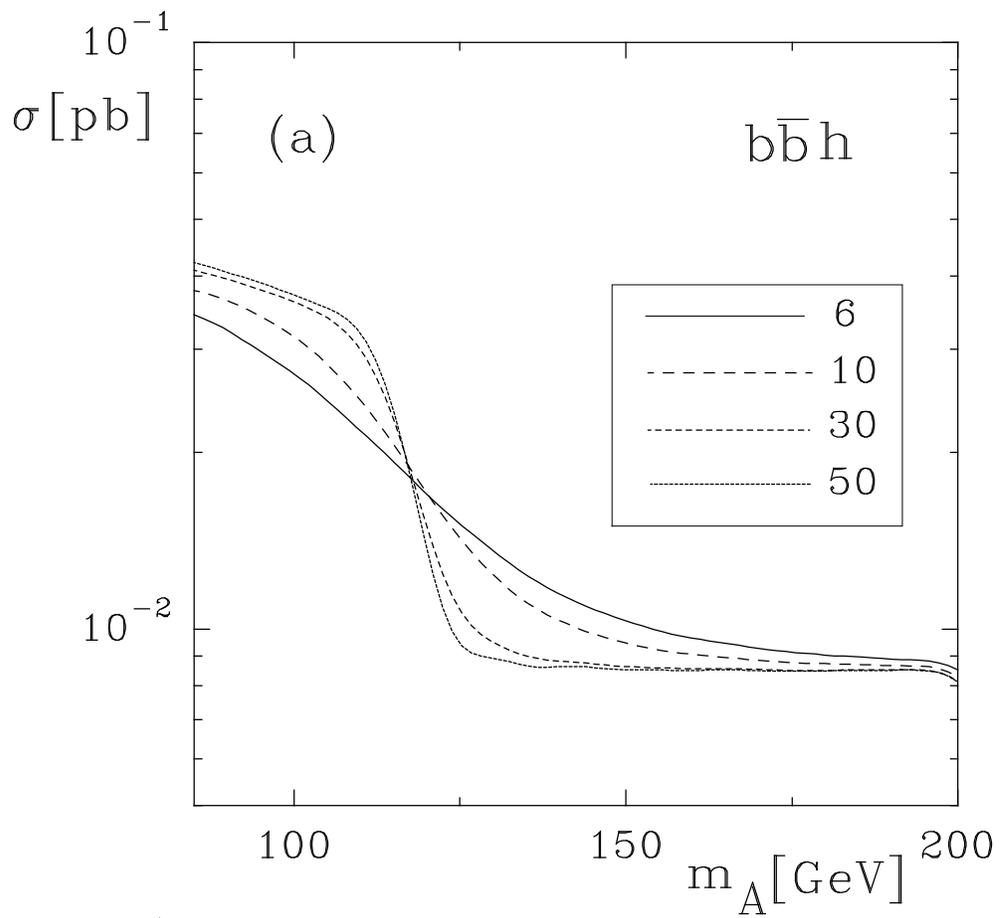
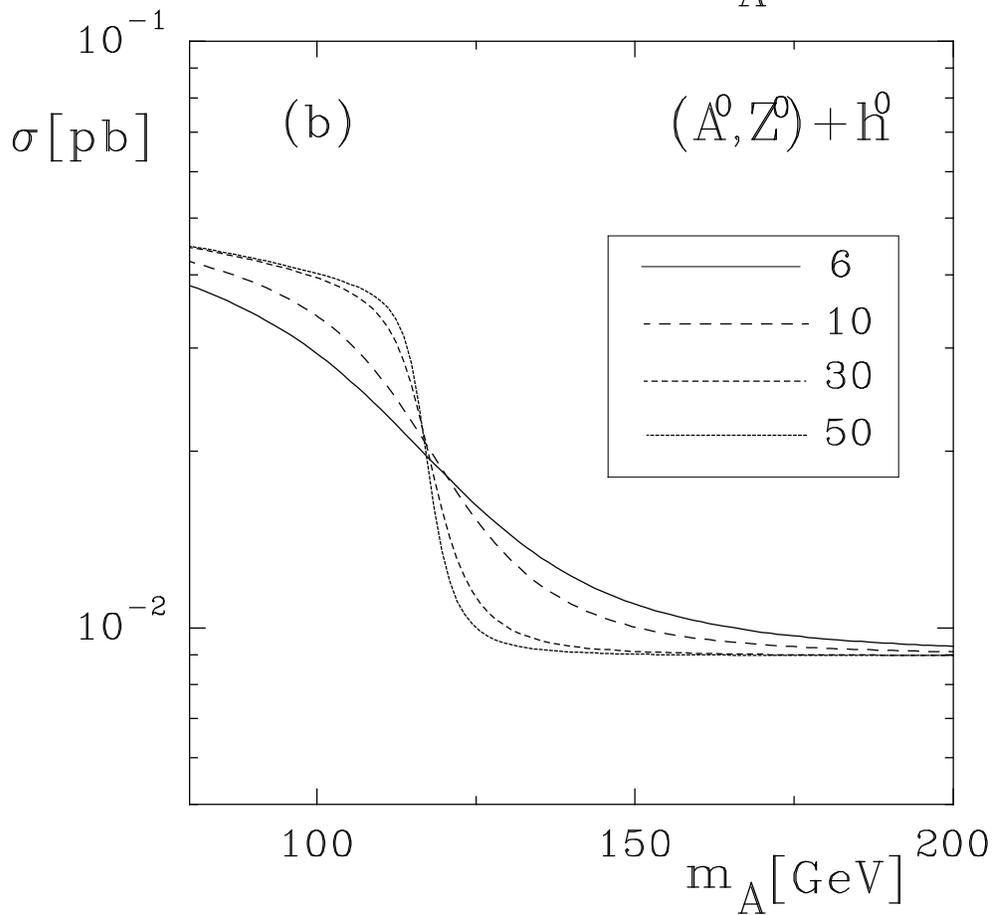
\noindent {\bf Fig. 4} Total Higgs production cross sections
$e^{+}e^{-}\rightarrow b\bar b h^0$ and $e^{+}e^{-}\rightarrow (A^0, Z^0) + h^0$
with $\sqrt{s} = 500$ and 4 different values of $\tan\beta$, 6, 10, 30, 50.
We have taken $m_{t} = 175$ $GeV$ and $M_{\stackrel {\sim}t} = 500$ $GeV$ and neglected
squark mixing.

\bigskip

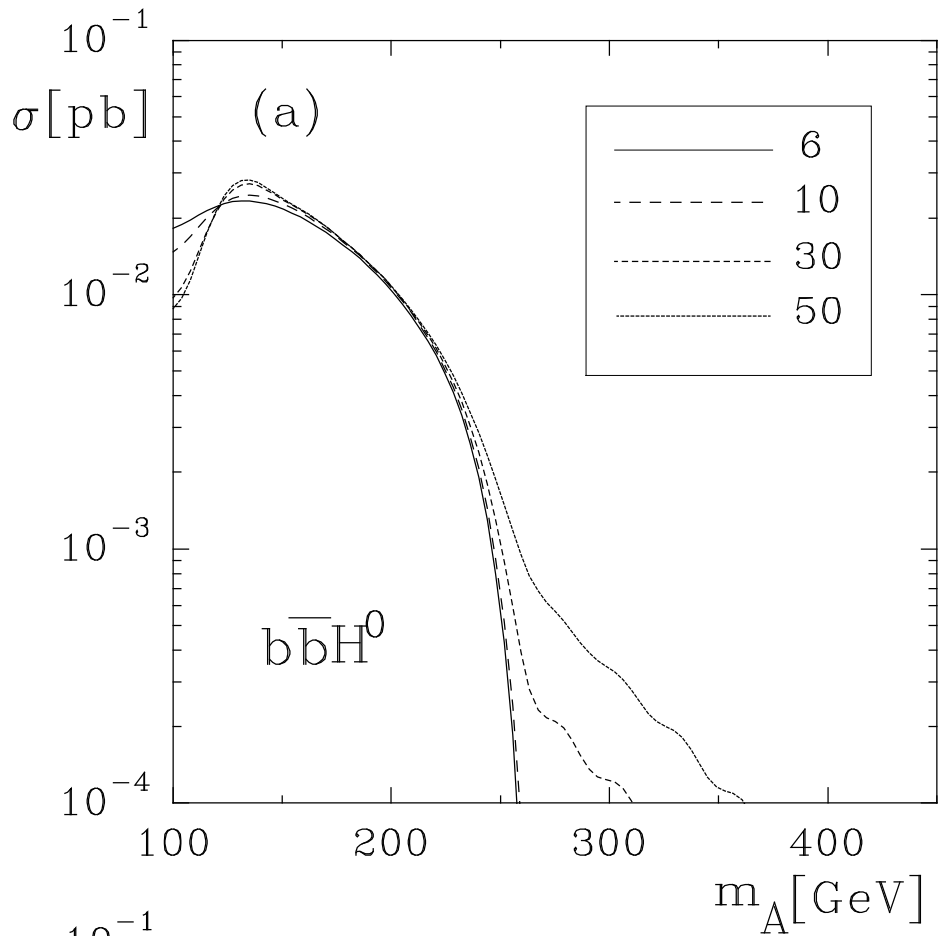
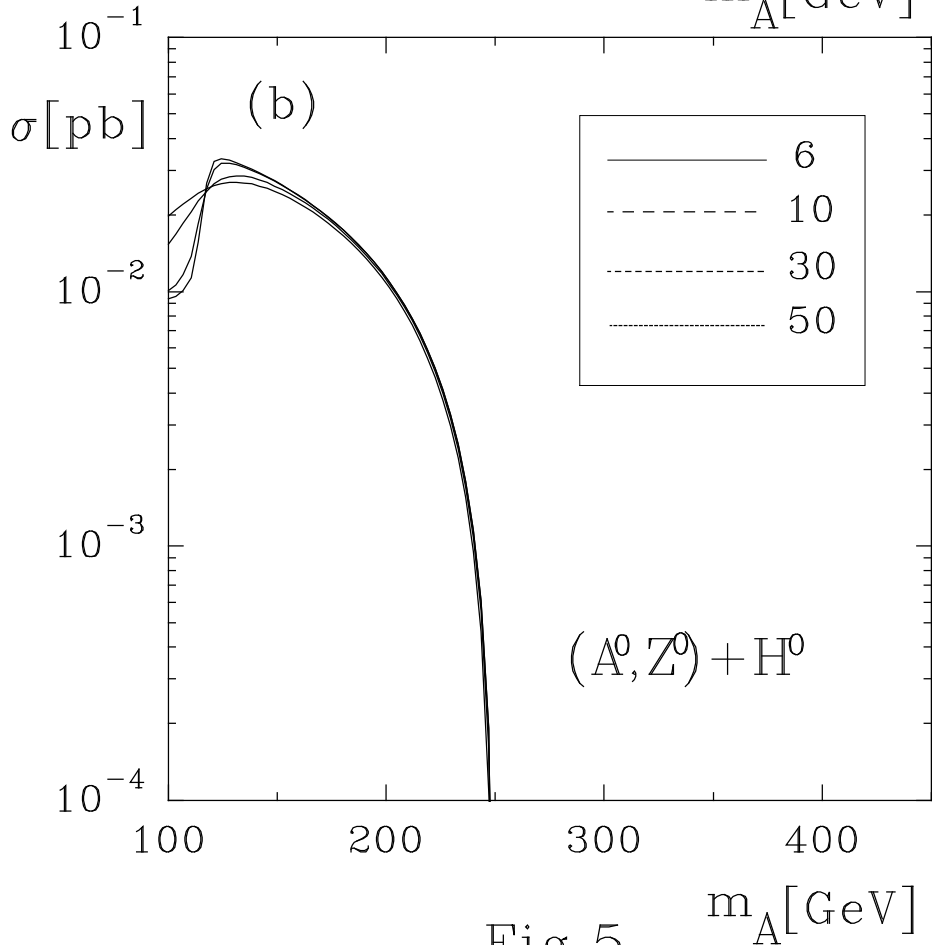
\noindent {\bf Fig. 5} Same as in Fig. 4, but for
$e^{+}e^{-}\rightarrow b\bar b H^0$ and $e^{+}e^{-}\rightarrow (A^0, Z^0) + H^0$.

\bigskip

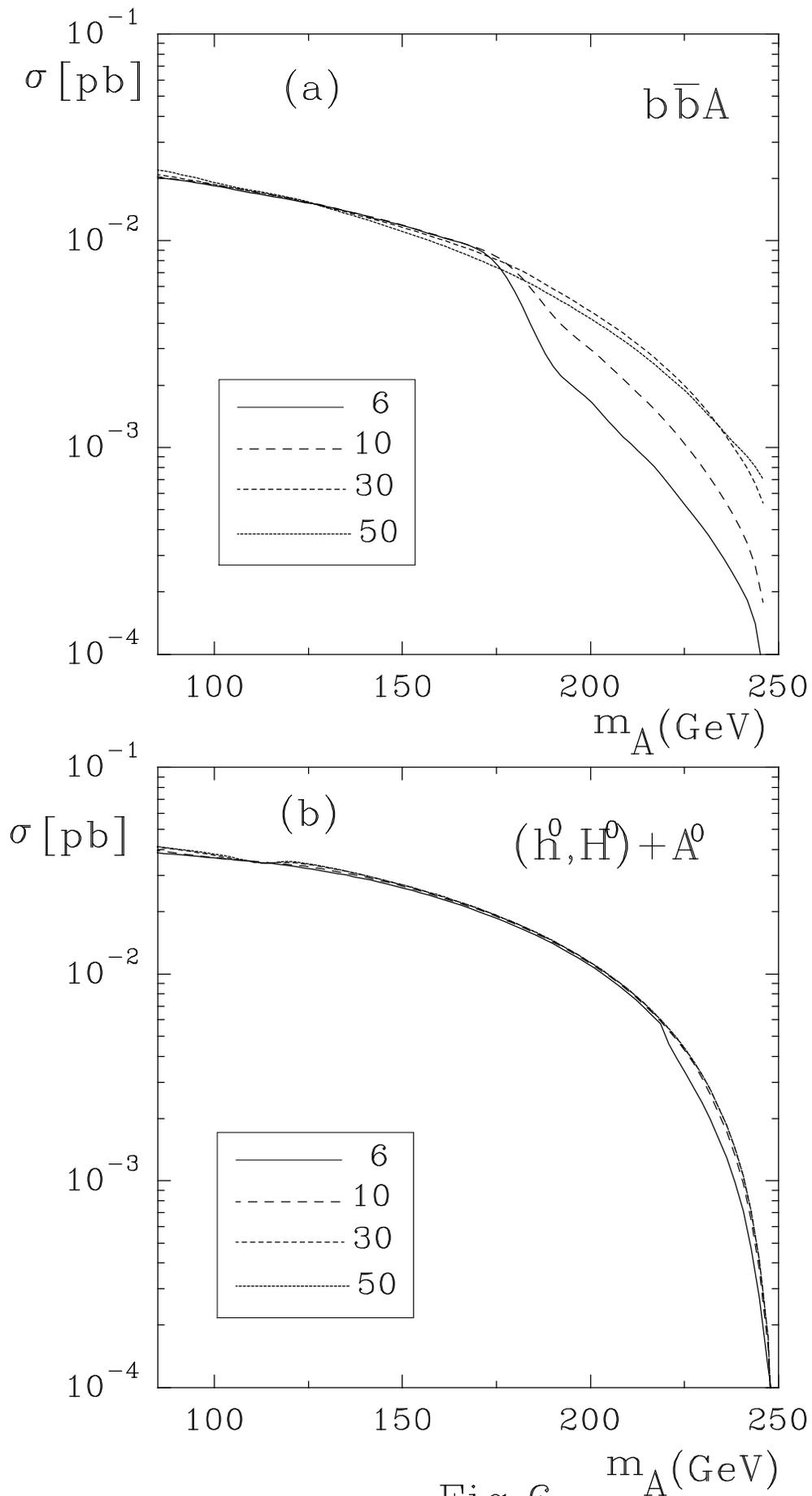
\noindent {\bf Fig. 6} Same as in Fig. 4, but for
$e^{+}e^{-}\rightarrow b\bar b A^0$ and $e^{+}e^{-}\rightarrow (h^0, H^0) + A^0$.

\bigskip

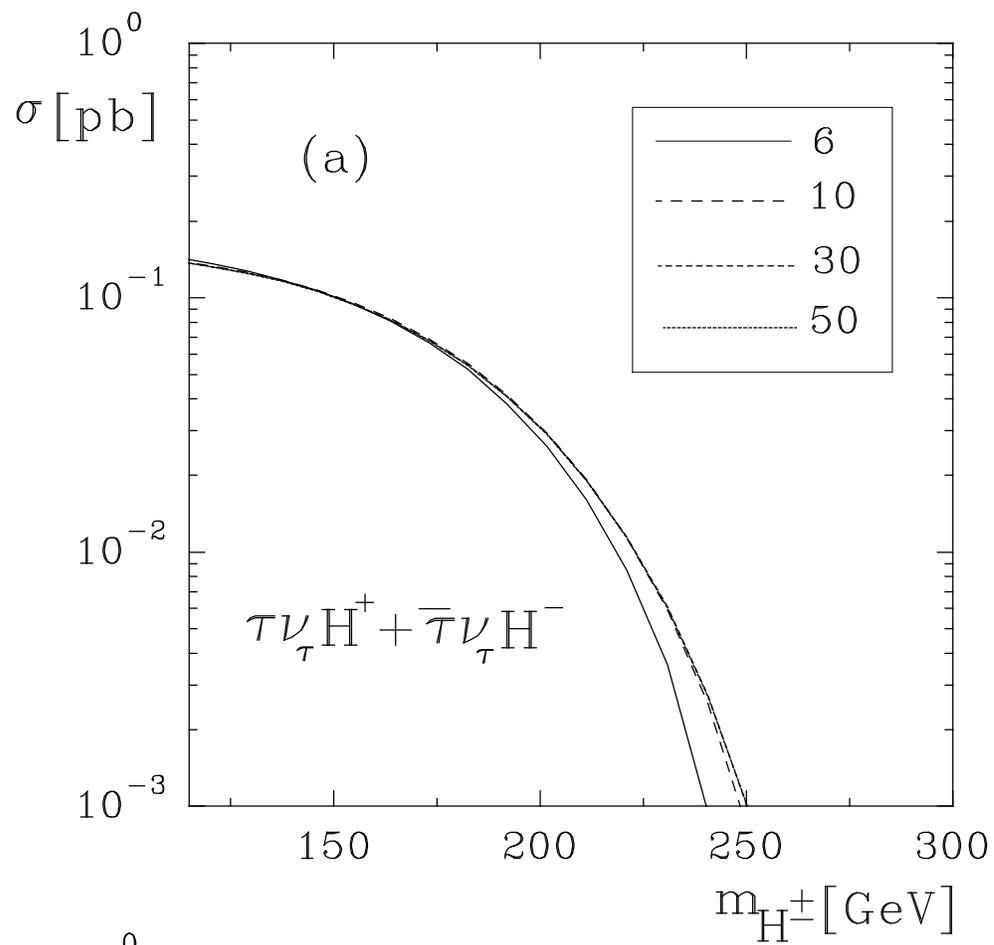
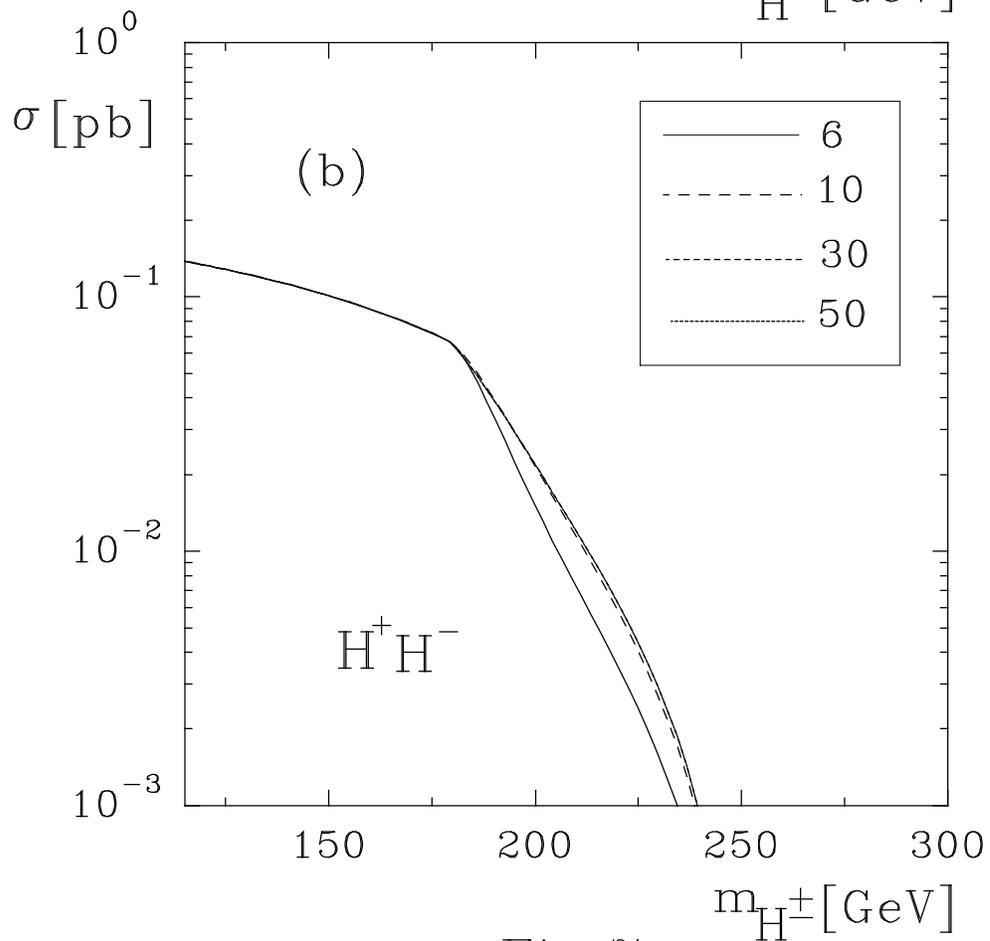
\noindent {\bf Fig. 7} Same as in Fig. 4, but for
$e^+e^-\to \tau^-\bar \nu_\tau H^+, \tau^+ \nu_\tau H^-$
and $e^{+}e^{-}\rightarrow H^+ H^-$.

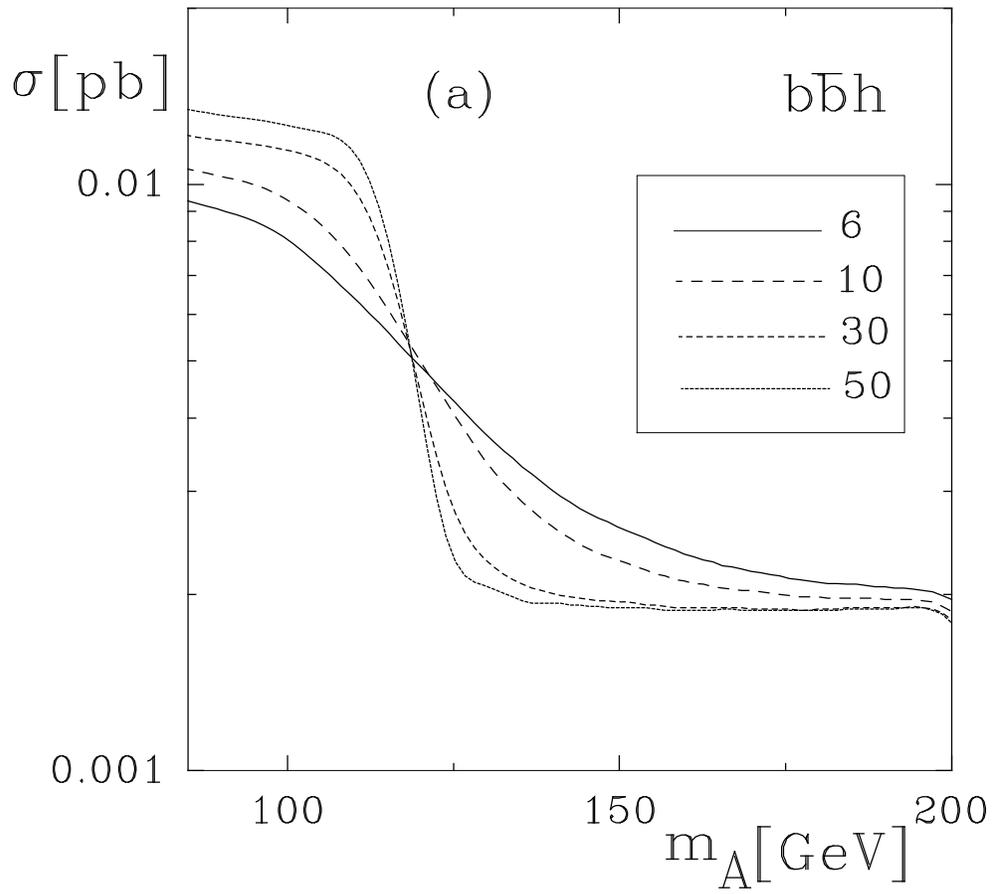
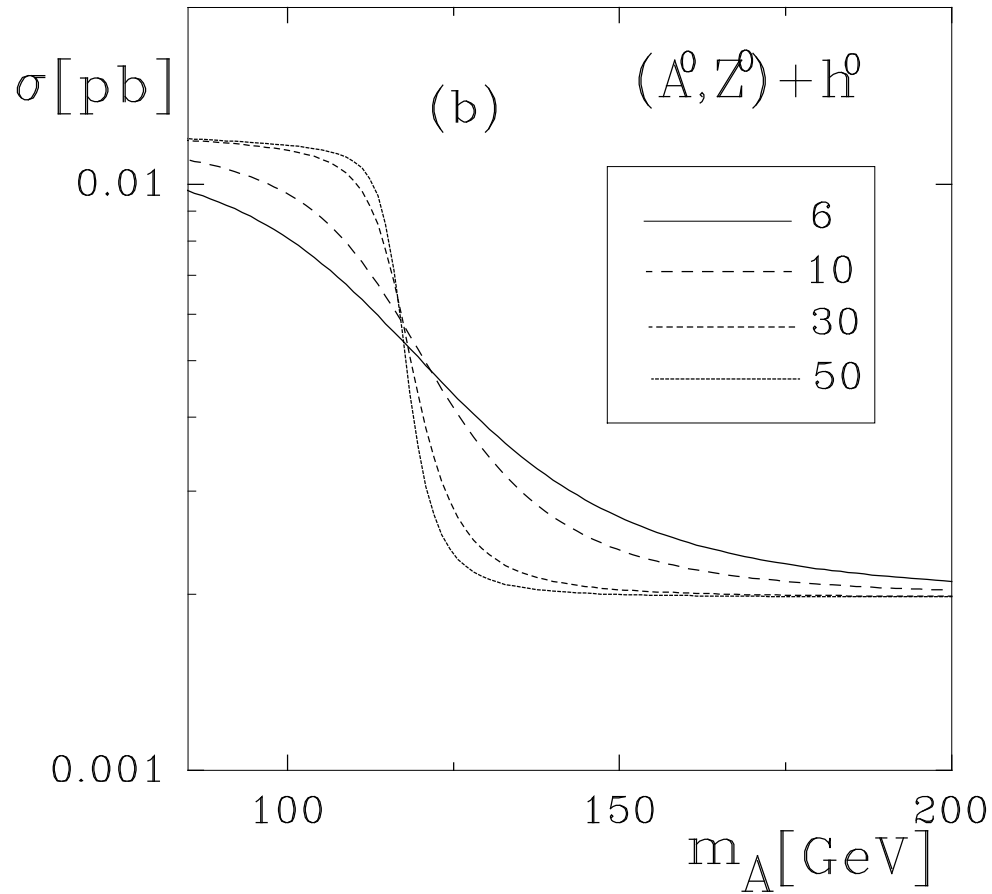
\noindent {\bf Fig. 8} Total Higgs production cross sections
for $\sqrt{s}= 1$ $TeV$ and 4 different values of $\tan\beta$, 6, 10, 30, 50.
We have taken $m_{t} = 175$ $GeV$, $M_{\stackrel {\sim} t} = 500$ $GeV$ and neglected
squark mixing. We display contours for $e^{+}e^{-}\rightarrow b\bar b h^0$ and
$e^{+}e^{-}\rightarrow (A^0, Z^0) + h^0$, in the parameters space
$(\sigma_T-m_{A^0})$.

\bigskip

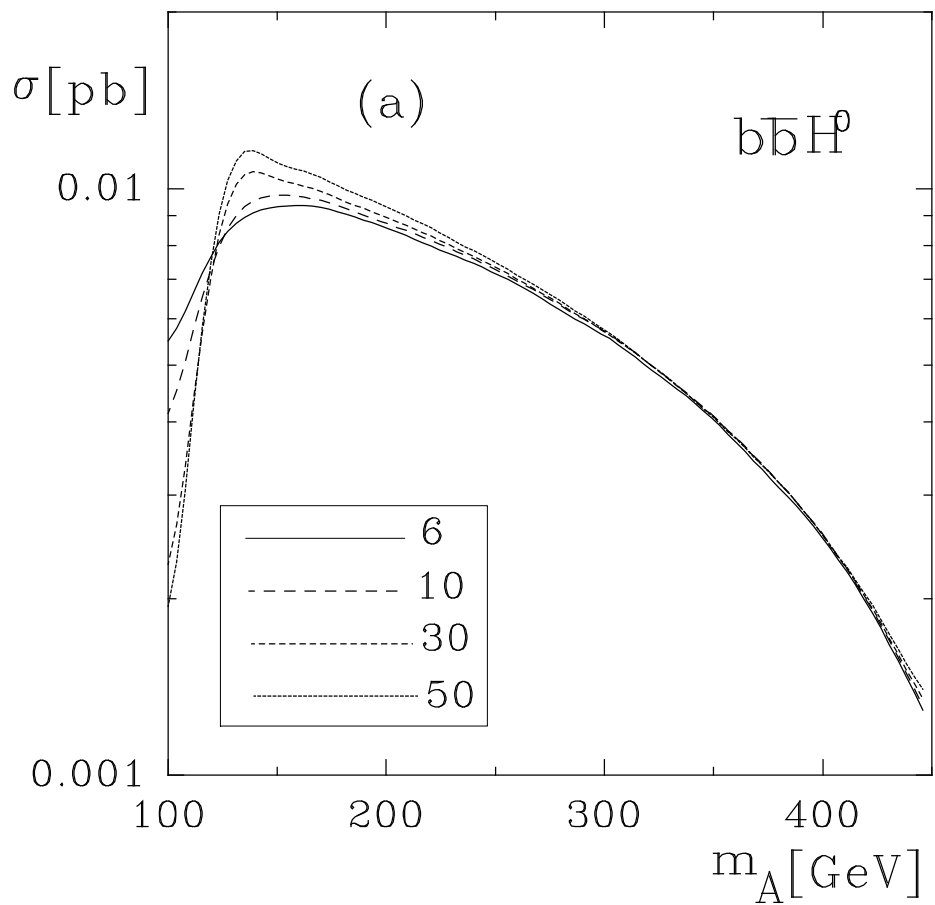
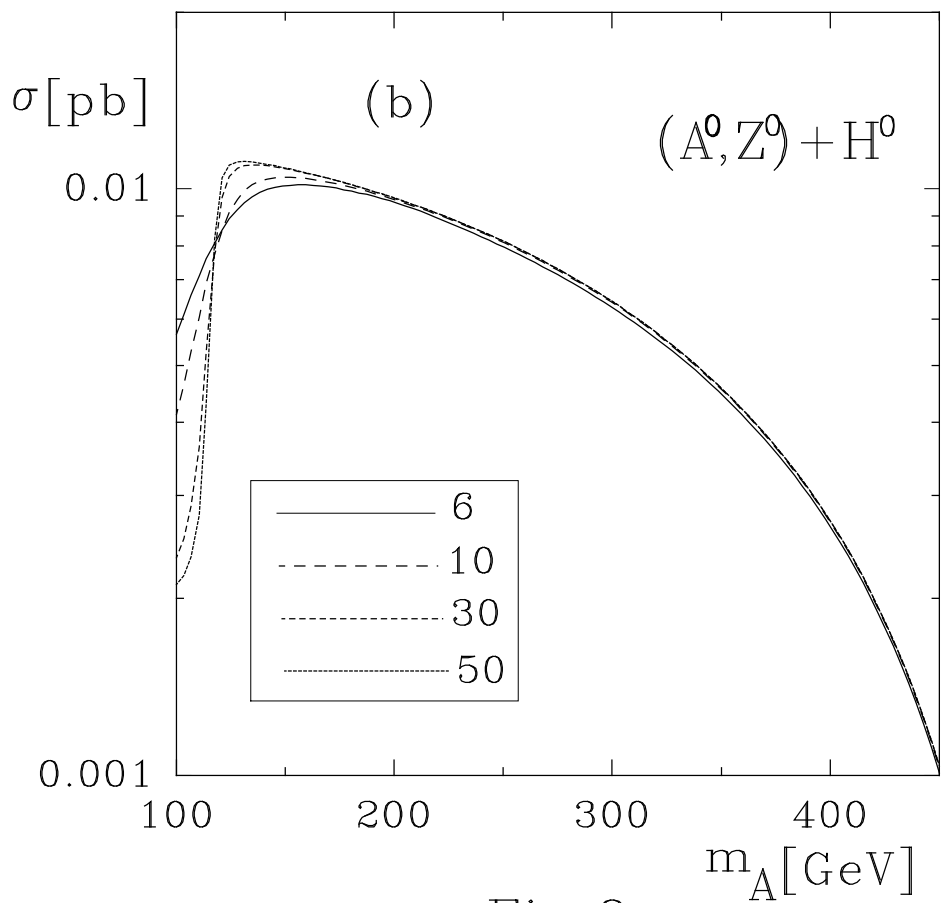
\noindent {\bf Fig. 9} Same as in Fig. 8, but for
$e^{+}e^{-}\rightarrow b\bar b H^0$ and $e^{+}e^{-}\rightarrow (A^0, Z^0) + H^0$.

\bigskip

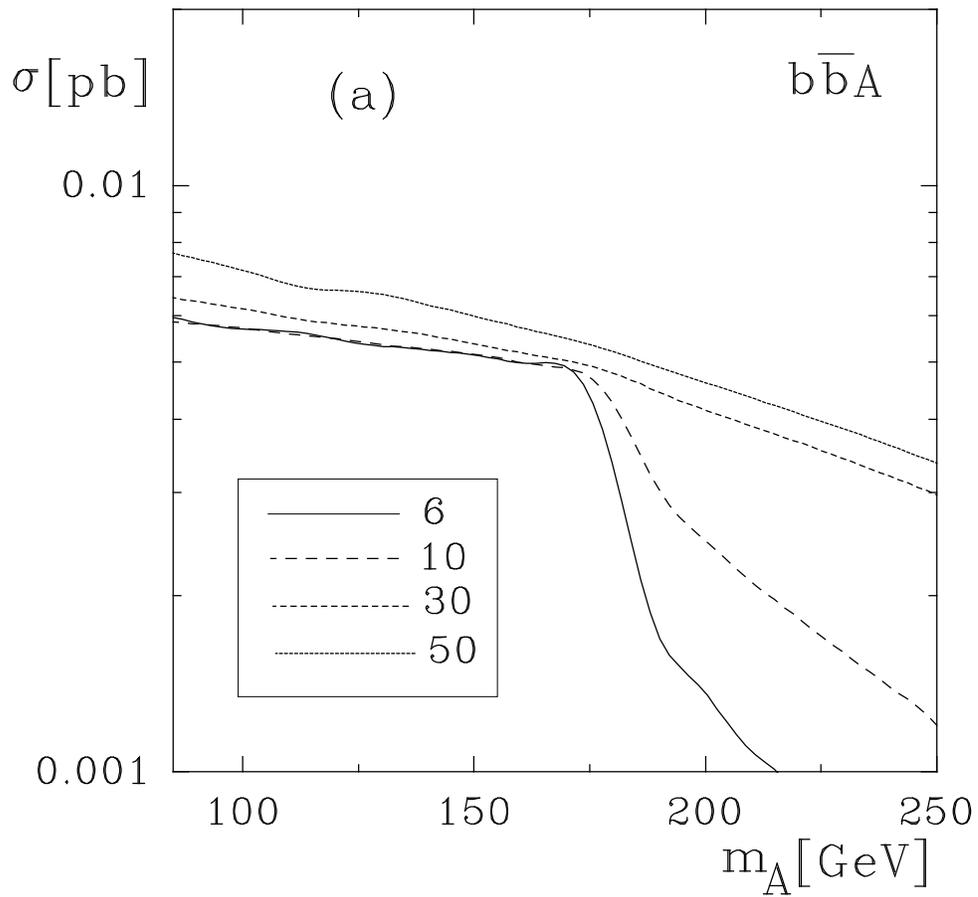
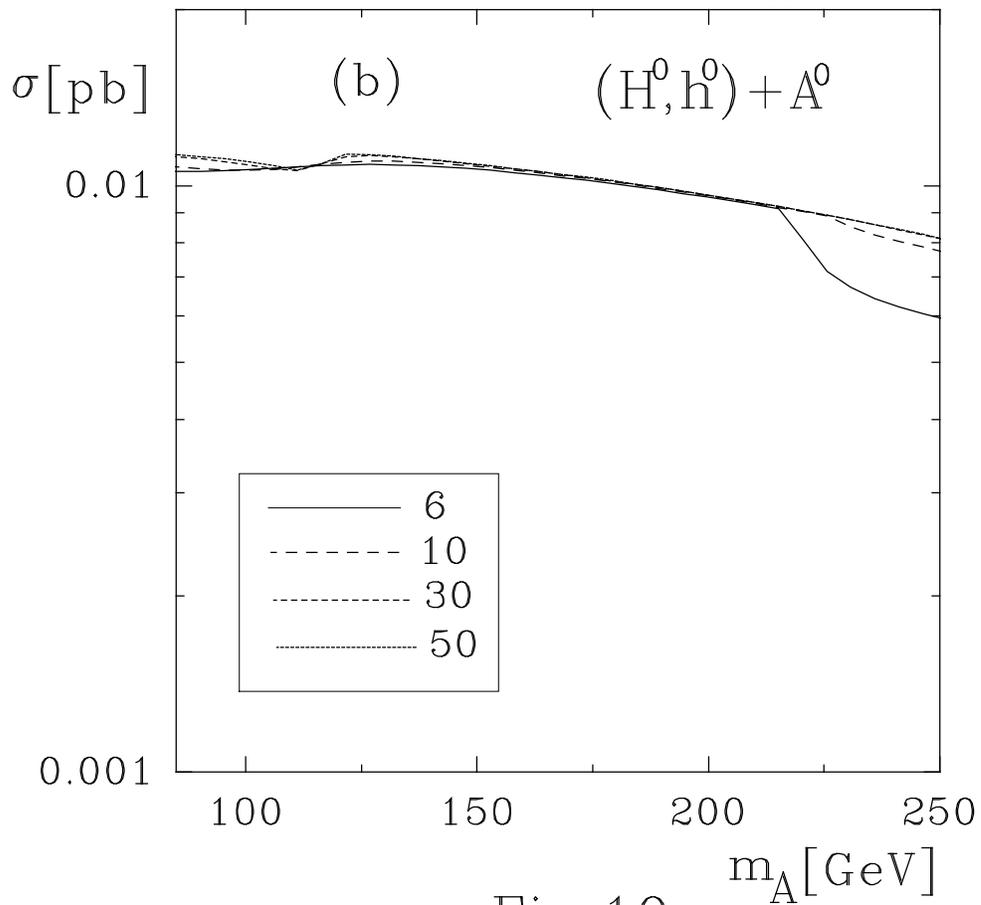
\noindent {\bf Fig. 10} Same as in Fig. 8, but for
$e^{+}e^{-}\rightarrow b \bar b A^0$ and $e^{+}e^{-}\rightarrow (h^0, H^0) + A^0$.

\bigskip

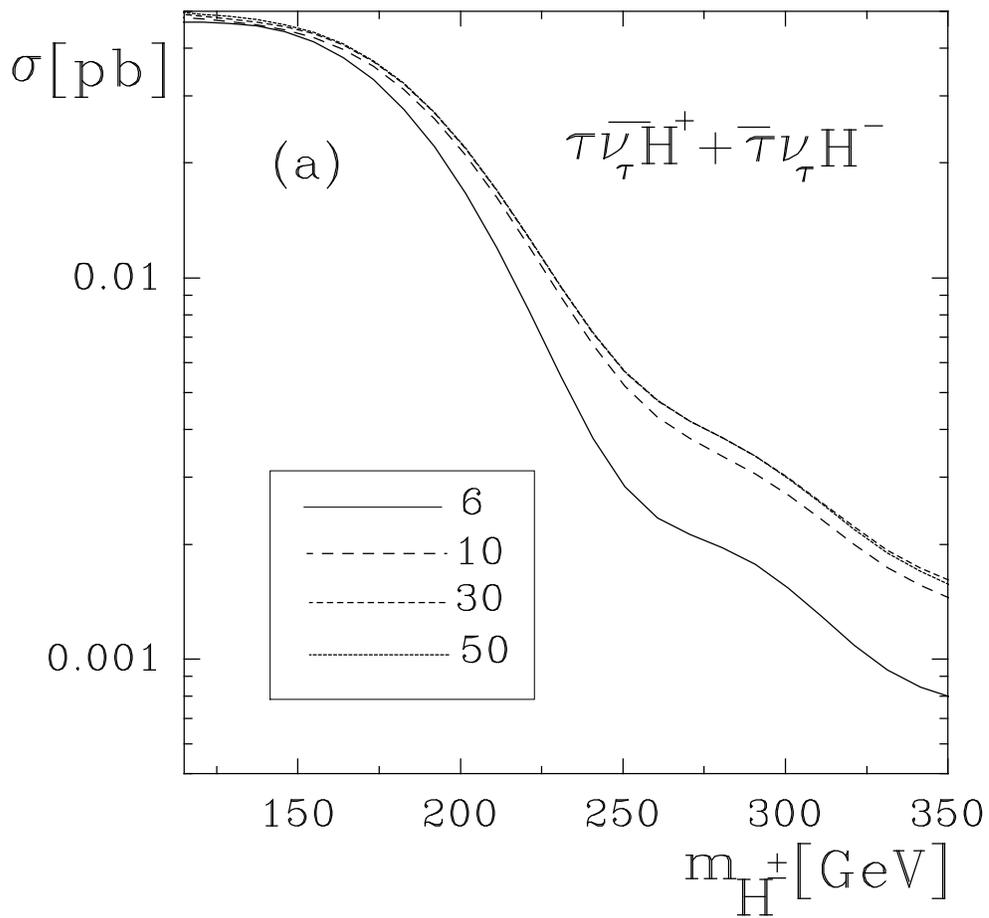
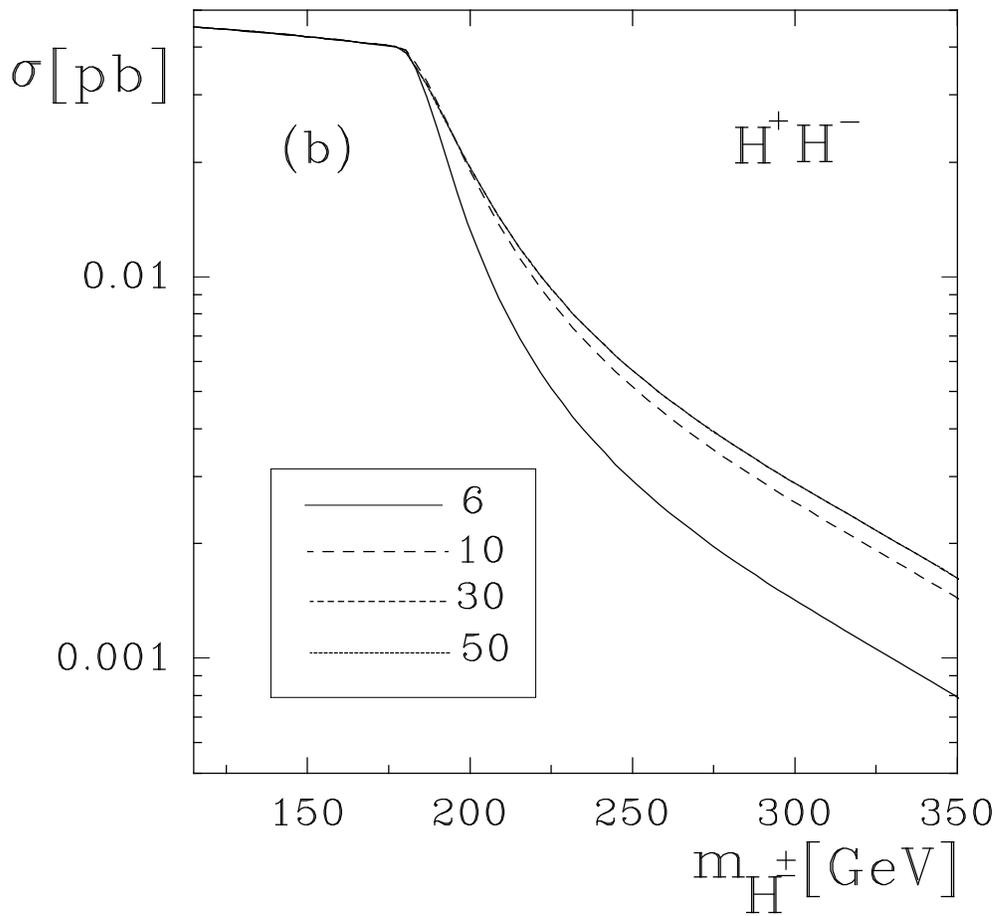
\noindent {\bf Fig. 11} Same as in Fig. 8, but for
$e^+e^-\to \tau^-\bar \nu_\tau H^+, \tau^+ \nu_\tau H^-$ and
$e^{+}e^{-}\rightarrow H^+ H^-$.

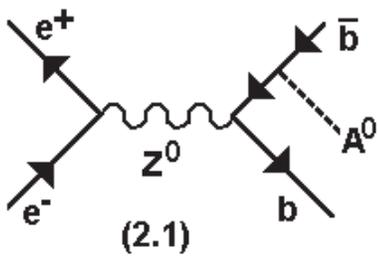
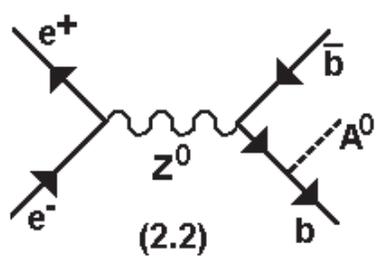
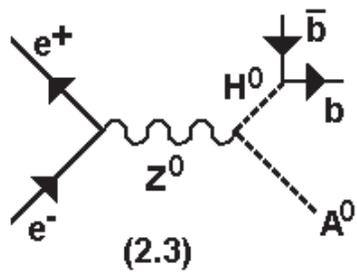
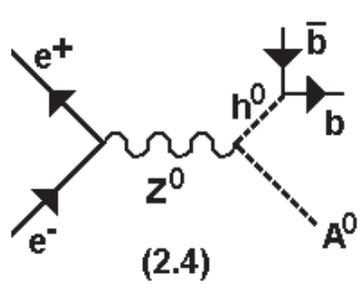
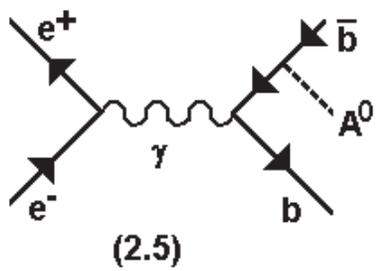
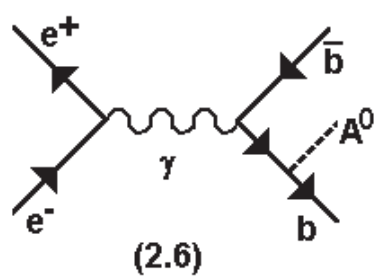
\end{document}